\documentclass[twocolumn]{aastex631}
\usepackage{array,multirow,graphicx,amsmath,amssymb}

\begin{document}

\title{Strongly magnetized accretion with low angular momentum produces a weak jet}

\correspondingauthor{Alisa Galishnikova}
\email{agalishnikova@flatironinstitute.org}

\author{Alisa Galishnikova}
\affiliation{Center for Computational Astrophysics, Flatiron Institute, 162 Fifth Avenue, New York, NY 10010, USA}
\affiliation{Department of Astrophysical Sciences, Princeton University, 4 Ivy Lane, Princeton, NJ 08544, USA}

\author{Alexander Philippov}
\affiliation{Department of Physics, University of Maryland, College Park, MD 20742, USA}

\author{Eliot Quataert}
\affiliation{Department of Astrophysical Sciences, Princeton University, 4 Ivy Lane, Princeton, NJ 08544, USA}

\author{Koushik Chatterjee}
\affiliation{Department of Physics, University of Maryland, College Park, MD 20742, USA}

\author{Matthew Liska}
\affiliation{Center for Relativistic Astrophysics, Georgia Institute of Technology, Howey Physics Bldg, 837 State St NW, Atlanta, GA 30332, USA}
\affiliation{Institute for Theory and Computation, Harvard University, 60 Garden Street, Cambridge, MA 02138, USA}
\begin{abstract}
We study the spherical accretion of magnetized plasma with low angular momentum onto a supermassive black hole, utilizing global general relativistic magnetohydrodynamic simulations. Black hole-driven feedback in the form of magnetic eruptions and jets triggers magnetized turbulence in the surrounding medium. We find that when the Bondi radius exceeds a certain value relative to the black hole's gravitational radius, this turbulence restricts the subsequent inflow of magnetic flux, strongly suppressing the strength of the jet. Consequently, magnetically arrested disks and powerful jets are not a generic outcome of the accretion of magnetized plasma, even if there is an abundance of magnetic flux available in the system. However, if there is significant angular momentum in the inflowing gas, the eruption-driven turbulence is suppressed (sheared out), allowing for the presence of a powerful jet.  Both the initially rotating and nonrotating flows go through periods of low and high gas angular momentum, showing that the angular momentum content of the inflowing gas is not just a feature of the ambient medium, but is strongly modified by the eruption and jet-driven black hole feedback. In the lower-angular-momentum states, our results predict that there should be dynamically strong magnetic fields on horizon scales, but no powerful jet; this state may be consistent with Sgr A* in the Galactic center.
\end{abstract}


\section{Introduction} \label{sec:intro}
General relativistic magnetohydrodynamic (GRMHD) simulations are frequently used to study accretion flows around supermassive black holes (BHs) \citep[e.g.,][]{EHTcode}. 
In the context of low-luminosity accretion flows, radiation is dynamically unimportant, which leaves the ideal GRMHD method scale-free. Typically, these models are initialized with a rotating Fishbone-Moncrief torus \citep{1976ApJtorus}, which is threaded with a dynamically weak magnetic field organized on large scales. In the absence of magnetic fields, this represents an equilibrium solution. 

The dynamics of magnetized accretion flows typically fall into two categories: Magnetically Arrested Disk (MAD) and Standard and Normal Evolution (SANE) \citep{Narayan2012MNRAS.426.3241N}. The MAD state is characterized by strong coherent magnetic fields that become dynamically important, effectively regulating the accretion flow by creating magnetic forces strong enough to counterbalance the gravitational pull of the BH. In particular, the magnetic flux threading the BH's horizon, $\Phi_{\rm BH}$, is argued to saturate at a maximum value for a given mass accretion rate, $\dot M_{\rm BH}$. The MAD flux is found to be approximately $50 \sqrt{\dot M_{\rm BH} r_g^2 c}$, where $r_g$ is the gravitational radius of the BH and $c$ is the speed of light. When the magnetic flux at the horizon exceeds the MAD limit, some of the flux erupts and escapes from the BH through magnetic reconnection \citep[][]{Ripperda2022}, leading to episodic fluctuations in $\Phi_{\rm BH}$. Most importantly, the MAD state is associated with the production of relativistic jets with electromagnetic luminosity exceeding the accretion power, $\dot M_{\rm BH} c^2$ \citep{Tchekhovskoy2011}.

The question remains whether the MAD state, observed in simulations initialized with an idealized rotating torus, is reached in realistic accretion scenarios. Recent global MHD simulations provide evidence that it can \citep{Ressler2023MNRAS.521.4277R}. Considering Wolf-Rayet stellar winds as a source of gas in the case of Sgr A* \citep{Ressler2020ApJ...896L...6R}, these studies demonstrated that the accretion flow can transition into a MAD state, even when the field is weak far from the BH. 

\cite{Ressler2021MNRAS} showed that a MAD could also form as a result of a low-angular-momentum magnetized flow onto a spinning BH -- an idealization of the stellar wind accretion problem. They examined various orientations of the initial magnetic field and found that the dynamics of the flow near the horizon closely resemble MAD dynamics. However, on larger scales, they argued that the jet is unstable due to the current-driven kink instability. A similar problem was investigated by \cite{Lalakos2024ApJ}. By increasing the Bondi radius by a factor of 10 compared to \cite{Ressler2021MNRAS}, and focusing on a magnetic field aligned with the BH's spin, they also found that the system quickly transitions to a MAD state. However, as time progressed, their simulation transitioned into a state without a powerful jet, which they argued to be a result of the jet becoming kink-unstable (by criteria described in \citealt{Tchekhovskoy2011,Bromberg2016MNRAS}) and disrupting the accretion flow. Thus, two different regimes, resulting from different Bondi radii, were identified.

Studying a similar problem, \cite{Kwan2023ApJl} conducted a series of GRMHD simulations of spherical accretion of magnetized medium with varying initial angular momentum. In contrast to \cite{Ressler2021MNRAS} and \cite{Lalakos2024ApJ}, they argued that no persistent jet can form in the zero-angular-momentum case. The inclusion of angular momentum resulted in a powerful jet capable of propagating to large distances. They attributed this to the dependence of the efficiency of the angular momentum transport on the magnetorotational instability.

The diversity and seeming incompatibility of the results in the literature on the properties of jets in low-angular-momentum flows motivates us to reexamine this problem. Spherical accretion of magnetized plasma is also the simplest problem in which to understand magnetic flux accumulation by a central black hole, a key problem in accretion theory given the central role of magnetic fields in angular momentum transport and jet/outflow production. 
The most idealized version of this scenario involves only a few key parameters: the spin of the BH, the Bondi radius of the surrounding medium, the strength of the magnetic field at the Bondi radius (characterized by plasma-$\beta$) and its orientation relative to the BH spin, and the initial angular momentum (or, equivalently, circularization radius) and its direction. In this study, we restrict our analysis to the case of aligned magnetic, rotation, and BH spin directions but study the effects of different Bondi radii, initial magnetic field strengths, black hole spin, and initial gas angular momentum.

The remainder of the paper is organized as follows. We describe the setup of the GRMHD simulations used in this work in \S~\ref{sec:method} and present our results in \S~\ref{sec:results}. We first provide an illustrative example of the dependence of the magnetized spherical accretion problem on the chosen Bondi radius in \S~\ref{sec:main}. Then, we discuss the dependence of our results on the key parameters of our study in \S~\ref{sec:parameters}; we consider the case of finite initial angular momentum -- and the time evolution of the resulting angular momentum -- in \S~\ref{sec:mom}.

\section{Method}\label{sec:method}
We conduct GRMHD simulations utilizing the GPU-accelerated code H-AMR \citep{liska2019hamr}, which solves the MHD equations in a conservative form in curved space-time. Employing horizon-penetrating Kerr-Schild (KS) coordinates $r_{KS}$, $\theta_{KS}$, $\phi_{KS}$, we concentrate on a nonspinning BH characterized by a dimensionless spin parameter $a=0$, as well as a highly spinning BH with $a=0.95$. In the case of a highly spinning BH, its spin axis aligns with the polar axis. Our results, however, are insensitive to the inclination of the metric relative to the polar singularity of the coordinate system\footnote{We demonstrate the insensitivity to the orientation with respect to the pole by tilting the BH spin by $90^{\circ}$ in Appendix~\ref{ap:convergence}.}.

We conduct simulations at three resolutions, which we label as small, medium, and large. These correspond to effective grid sizes, $N_r \times N_\theta \times N_\phi$, of $384 \times 144 \times 256$, $812 \times 384 \times 512$, and $1536 \times 768 \times 1024$ cells. We employ ``derefinement'' near the pole, which efficiently reduces computational costs while maintaining a comparable spatial resolution near both the pole and the equator. The outer boundary, $R_{\rm out}$, is set to $10^4 r_g$ in all simulations unless stated otherwise (the outer boundary of one of the simulations was extended to $10^5r_g$ to accommodate a Bondi radius of $5000r_g$). Our primary focus will be on the medium-resolution runs, as we observe the convergence of the main results with resolution (see Appendix~\ref{ap:convergence}). Throughout the paper, we measure distance and time in $r_g=GM/c^2$ and $r_g/c$ respectively, where $G$ is the gravitational constant and $M$ is the mass of the central BH.

We build upon previous work on spherical accretion with a large-scale poloidal magnetic field by exploring a broader parameter space. We consider values of the Bondi radius $r_b$ of $100r_g$, $500r_g$, $1000r_g$, and $5000r_g$:
\begin{equation}
    r_b = \frac{2M}{c_s^2},
\end{equation}
where $c_s$ represents the initial gas sound speed. Initially, a zero-angular-momentum fluid with $\gamma=5/3$ is distributed throughout the simulation domain, characterized by a uniform density $\rho_0$ (we set $\rho_0=1$ in code units). The internal energy density is also uniform and determined by $r_b$:
\begin{equation}
    u_g = \frac{2 \rho_0 M }{\gamma (\gamma-1) r_b}.
\end{equation}
The initial plasma-$\beta$, defined as the ratio of thermal pressure $P_{\rm th}$ to magnetic pressure $P_B$, sets the magnetic field magnitude $\sqrt{b_0^2}$ via
\begin{equation}
    \beta_0 = (\gamma-1) \frac{2 u_g }{b_0^2}.
\end{equation}
Throughout the paper, $b^\mu$ represents the fluid-frame magnetic field, while $B^i$ is measured in the lab frame. Similar to \cite{Lalakos2024ApJ}, we introduce $2\%$ random noise to the initial energy density or plasma density to break the symmetry in the simulation domain.

The initial magnetic field in the laboratory frame $B^i$ is strictly poloidal, aligned with the spin axis of the BH, and determined by a vector potential $A_\phi$. Following a similar approach as outlined in \citet{Ressler2021MNRAS}, the strength of the initial magnetic field (as well as the plasma density) is suppressed close to the BH, i.e., within $r < 6r_g$: 
\begin{equation}
    A_{\phi} = \frac{1}{2} b_0 r^2 \sin^2 \theta \times 
    \begin{cases}
        e^{5 \left( 1 - 6/r\right)}, \ r < 6 r_g\\
        1, \ {\rm otherwise} .
    \end{cases}
\end{equation}
Outside of $6r_g$, plasma-$\beta$ is initialized as uniform.

We set $\beta_0=10^2$ in the initial conditions of our $r_b=100r_g$ simulation. In a magnetized Bondi solution with a radial monopole magnetic field, $B^r$ undergoes stretching, $B^r \propto r^{-2}$. Therefore, to maintain similarity in the plasma-$\beta$ and magnetic flux ($\Phi_{\rm BH}$, described below) near the BH across different runs, we scale it according to the Bondi profile, $\beta \propto r^{3/2}$. Thus, we set $\beta_0=10^3$ for $r_b=500r_g$ and $\beta_0 = 3 \times 10^3$ for $r_b=1000r_g$. This choice is not critical, however, as our results are in fact insensitive to the specific $\beta_0$ value, as long as it is initially above unity. We validate this statement by considering 4 values of $\beta_0$ for $r_b=500r_g$ (see \S~\ref{sec:parameters}): $10$, $10^2$, $10^3$, and $10^4$. This encapsulates our simulations' three key parameters: spin $a$, Bondi radius $r_b$, and initial plasma-$\beta_0$. We also discuss the dependence of our results on the initial angular momentum of the gas, parameterized using the circularization radius, $r_c$, in \S~\ref{sec:mom}. 

The Bondi accretion rate for an adiabatic index of $\gamma = 5/3$ is:
\begin{equation}
    \dot{M_b} = \pi \rho r_b^2 c (r_b / r_g)^{-1/2} \propto r_b^{3/2},
\end{equation}
which sets the units for mass accretion rate measurements at the horizon:
\begin{equation}
    \dot{M} = -\int dA \rho u^r,
\end{equation}
where the integration is performed over a spherical shell, $dA = \sqrt{-g} d\theta d\phi$; $u^\mu$ is the contravariant velocity, $g$ represents the determinant of the Kerr-Schild metric. We also measure the total energy flux at the horizon as:
\begin{equation}
    \dot{E} = \int dA \left( \left[ \rho + \gamma u_g + b^2 \right] u^r u_t - \frac{b^r b_t}{4 \pi} \right),
\end{equation}
which provides an estimate of the outflow energy efficiency $\eta = \left( \dot{E} - \dot{M} \right) / \dot{M}$. A commonly used notation for the dimensionless magnetic flux through one hemisphere (hence the $1/2$ factor) is defined as:
\begin{equation}
    \phi_{\rm BH} = \frac{\Phi_{\rm BH} }{\sqrt{\dot{M} r_g^2 c }} =  \frac{\int dA |B^r| }{2\sqrt{\dot{M} r_g^2 c }}
\end{equation}
where $\Phi_{\rm BH} = 0.5 \int dA |B^r| $ is the unnormalized magnetic flux. In practice, we measure the above quantities at $5r_g$ to minimize uncertainties in measuring the mass accretion rate related to the chosen floor values\footnote{We have also examined all quantities at the horizon and found no significant difference in the time evolution of $\dot{E}$ and $\dot{M}$, as their average profiles remain relatively flat near the black hole.} {(except $\Phi_{\rm BH}$, which is measured at the horizon)}.

\section{Results}\label{sec:results}
First, we focus on the zero initial angular momentum case and the illustrative comparison of $r_b=100r_g$ and $r_b = 500r_g$ (with $\beta_0 = 10^2$ and $\beta_0=10^3$ respectively) at $a=0$ and $a=0.95$ in \S~\ref{sec:main}. These results are in fact more generic than they might seem because we find a convergence of our results as we increase $r_b$ yet further and vary $\beta_0$ (\S~\ref{sec:parameters}). Finally, the impact of angular momentum is discussed in \S~\ref{sec:mom}. 

\subsection{Early evolution}\label{sec:main}
In this section, we discuss the time evolution of our four main cases. We then discuss the role of eruption episodes as well as the magnetic field evolution. The nonspinning case converges with time and reaches a steady state. The highly spinning case initially reaches a similar steady state. However, at a later time, the dynamics of the spinning case with a larger Bondi radius, $r_b=500r_g$, is governed by the angular momentum generated by the BH feedback; this will be discussed in more detail in \S~\ref{sec:mom}.

\subsubsection{Temporal evolution}
Initially, the accreting plasma undergoes freefall onto the BH, bringing the magnetic flux toward the horizon. Consequently, with the increasing accretion rate at the horizon, plasma properties such as density and velocity conform to the expected Bondi profiles. As long as the magnetic field is dynamically subdominant, $B^\theta$ is advected by the radial velocity, while $B^r$ experiences stretching according to $B^r \propto r^{-2}$. Therefore, the total pressure for a Bondi-like flow profile is given by:
\begin{equation}
    P_{\rm tot}(r, \theta) = \frac{2\rho_0 M}{\gamma r_b} \left[ \left( \frac{r_b}{r} \right)^{5/2} + \frac{1}{\beta_0} \left[ \sin^2 \theta + \cos^2 \theta \left( \frac{r_b}{r}\right)^4 \right] \right],
    \label{eq:pressure}
\end{equation}
where the first and second terms correspond to the thermal and magnetic pressure components respectively. For $\beta_0 \gg 1$, this yields $P_B \ll P_{\rm th}$ close to the equatorial plane $\theta = \pi/2$. At higher altitudes, {close to the BH, $P_B$ exceeds $P_{\rm th}$ due to its stronger dependence on radius compared to $P_{\rm th}$}. 

\begin{figure*}
    \centering
    \includegraphics[width=\textwidth]{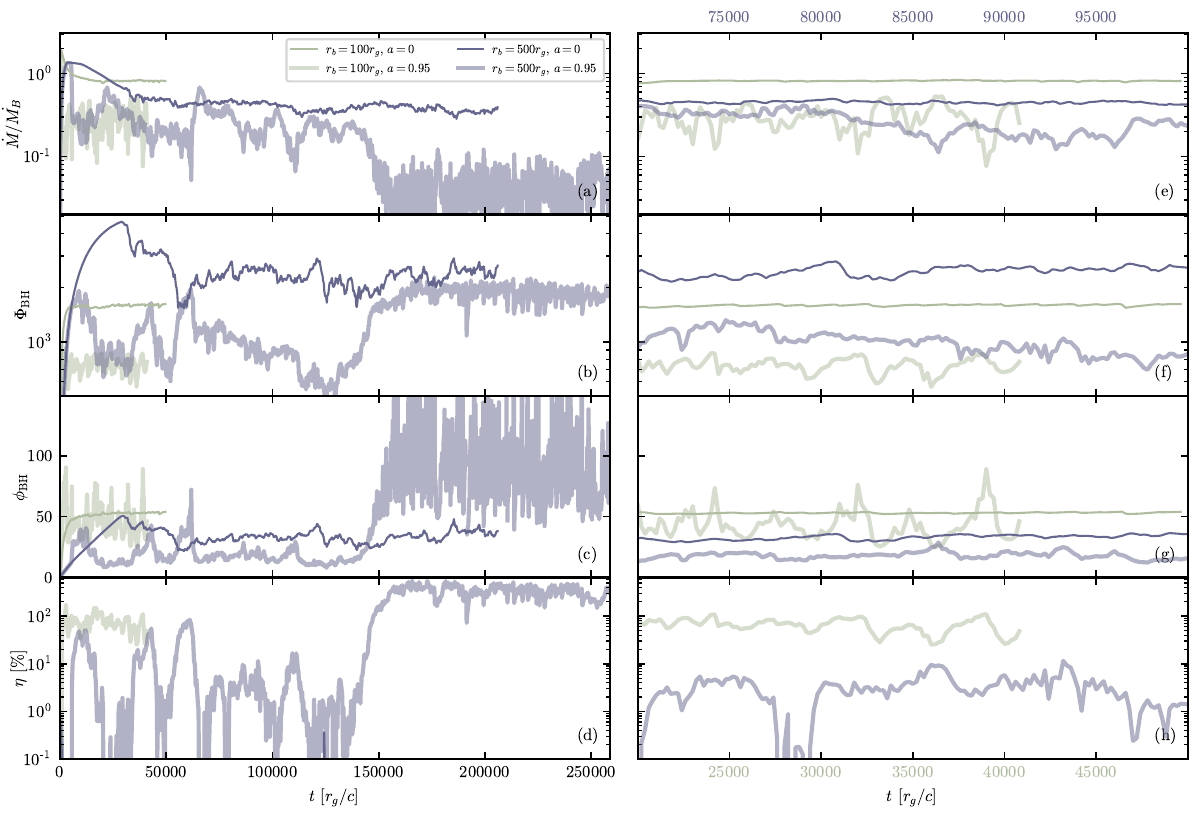}
    \caption{Comparison of the time evolution of the (a) accretion rate $\dot M/ \dot M_B$; (b) magnetic flux $\Phi_{\rm BH}$; (c) normalized dimensionless magnetic flux $\phi_{\rm BH}$, and (d) outflow energy efficiency $\eta$ for different $r_b$ ($100r_g$ and $500r_g$ are shown by different colors) and spin values ($a=0$ and $a=0.95$ are represented by thin and thick lines respectively). All quantities are measured at $5r_g$. On the right panel ((e)-(h)), a zoom-in of panels ((a)-(d)) is shown. Note the different x-axes for $r_b=100r_g$ (bottom x-axis) and $r_b=500r_g$ (top x-axis), represented by the corresponding colors.}
    \label{fig:time-evolution}
\end{figure*}
As the field becomes dynamically important, it starts influencing the accretion dynamics. Analogous to the MAD scenario, the accretion stalls, leading to the thinning of the inflow streams and the onset of magnetic reconnection. This induces a partial evacuation of the accretion flow, constraining both the accretion rate and magnetic flux at the horizon. This is evident in Figure~\ref{fig:time-evolution}, where we present the time evolution of the accretion rate $\dot{M}$ (first row), magnetic flux $\Phi_{\rm BH}$ (second row), dimensionless magnetic flux $\phi_{\rm BH}$ (third row), and the outflow energy efficiency $\eta$ (fourth row), measured at $5r_g$. Simulations with smaller, $r_b=100r_g$, and larger, $r_b=500r_g$, Bondi radii are shown by different colors; thin and thick lines represent $a=0$ and $a=0.95$, respectively. The total time evolution is shown in the left row. A zoom-in for short periods of time is shown in the right row, represented by two colors for $r_b=100r_g$ (bottom x-axis) and $r_b=500r_g$ (top x-axis); this shows the short-timescale variations more clearly. 

In all realizations, we see the presence of eruption episodes, which leads to the mass accretion rate $\dot{M}$ saturating just below $\dot{M}_{\rm B}$ (a). Smaller $r_b$ of $100r_g$ cases (green lines) show {an initial gradual increase of $\Phi_{\rm BH}$ from 0 and its subsequent saturation (b)}. In contrast, $r_b=500r_g$ cases show that $\Phi_{\rm BH}$ decreases after the initial eruption and saturates below its maximum value. 

At a smaller Bondi radius (green lines), the dimensionless magnetic flux reaches a value indicative of the MAD state, $\phi_{\rm BH}\approx 50$ (c). The energy outflow efficiency, $\eta$, is above $100\%$, demonstrating efficient jet production through the Blandford-Znajek mechanism for $a=0.95$ case (thick green line). {For $r_b=100r_g$ and $a=0.95$,} the oscillations in $\Phi_{\rm BH}$, $\phi_{\rm BH}$, and $\eta$ are similar to reconnection-powered eruptions found in MADs that are produced in simulations starting from rotating tori ((f)-(h)).

At a larger Bondi radius of $r_b=500r_g$, however, the quasi-state dimensionless magnetic flux is smaller than for $r_b=100r_g$. This trend persists for both nonspinning (thin lines) and spinning (thick lines) cases though the effect is stronger for the spinning BH (thick purple line saturates at a smaller value, compared to the thin purple line in panels (b) and (c)). 

Fig.~\ref{fig:time-evolution} shows that for $r_b=500r_g$, the BH is incapable of accreting magnetic flux beyond a certain threshold, despite the abundance of magnetic flux in the surrounding medium. For a highly spinning BH, this lack of significant inflow of magnetic flux {(indicated by $\Phi_{\rm BH}$ saturating below its maximum value)} results in the absence of a powerful jet; the outflow efficiency is only $\eta \sim 1 \% $. 

At a later time, the spinning BH with a larger Bondi radius (thick purple line) shows an increased $\phi_{\rm BH}$ and efficient energy outflow $\eta$ in the form of a powerful jet. We find such periodic jets to be a consequence of a periodic increase in the angular momentum in the inflow, which will be explored in more detail in \S~\ref{sec:mom}.

\begin{figure*}
    \centering
    \includegraphics[width=\textwidth]{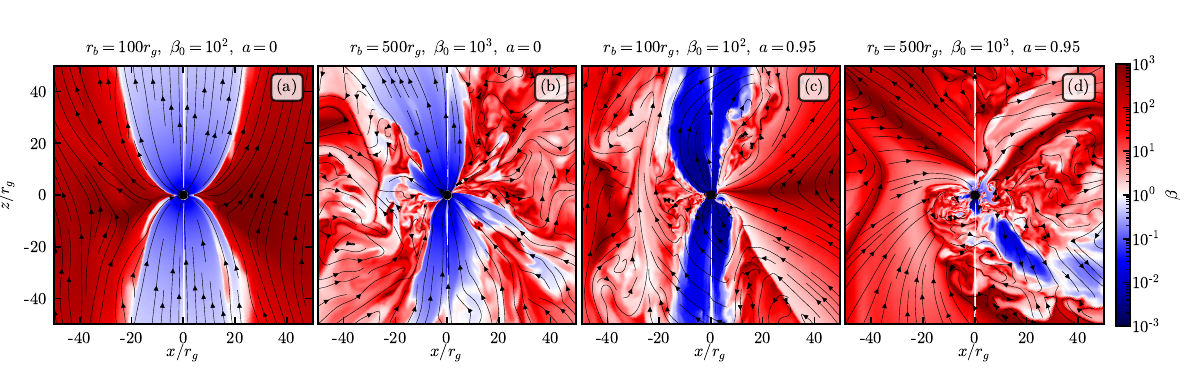}
    \caption{Plasma-$\beta$ in the poloidal plane for runs with different initial parameters $r_b$ and $a$; \textbf{black lines represent magnetic field lines}. The first two panels show $a=0$, and the last two show $a=0.95$. (a): $r_b=100r_g, \ \beta=10^2,\ a=0$; (b): $r_b=500r_g, \ \beta=10^3,\ a=0$; (c): $r_b=100r_g, \ \beta= 10^2,\ a=0.95$; (d): $r_b=500r_g, \ \beta=10^3, \ a=0.95$. The snapshots are taken at a time $40000r_g/c$ for $r_b=100r_g$ and $120000r_g/c$ for $r_b=500r_g$. }
    \label{fig:beta-side-by-size}
\end{figure*}
In Figure~\ref{fig:beta-side-by-size}, we present a side-by-side comparison of plasma-$\beta$ in the same four simulations in the poloidal $r-\theta$ plane with magnetic field lines shown by black lines. The first two columns illustrate the case of a nonspinning BH with $r_b=100r_g$ (a) and $r_b=500r_g$ (b). Panels (c) and (d) showcase a highly spinning BH, $a=0.95$, again with $r_b=100r_g$ and $r_b=500r_g$, respectively. In both cases where $r_b$ is smaller ($100r_g$, corresponding to higher gas temperature), a more laminar flow is observed compared to the scenario with $r_b=500r_g$. Moreover, in the case of a spinning BH, a smaller $r_b$ shows the presence of a jet (c), which is not seen in the $r_b=500r_g$ case (d). The snapshots are taken at the time of $40000r_g/c$ and $82000r_g/c$ for $r_b=100r_g$ and $r_b=500r_g$, respectively.

\begin{figure*}
    \centering
    \includegraphics[width=\textwidth]{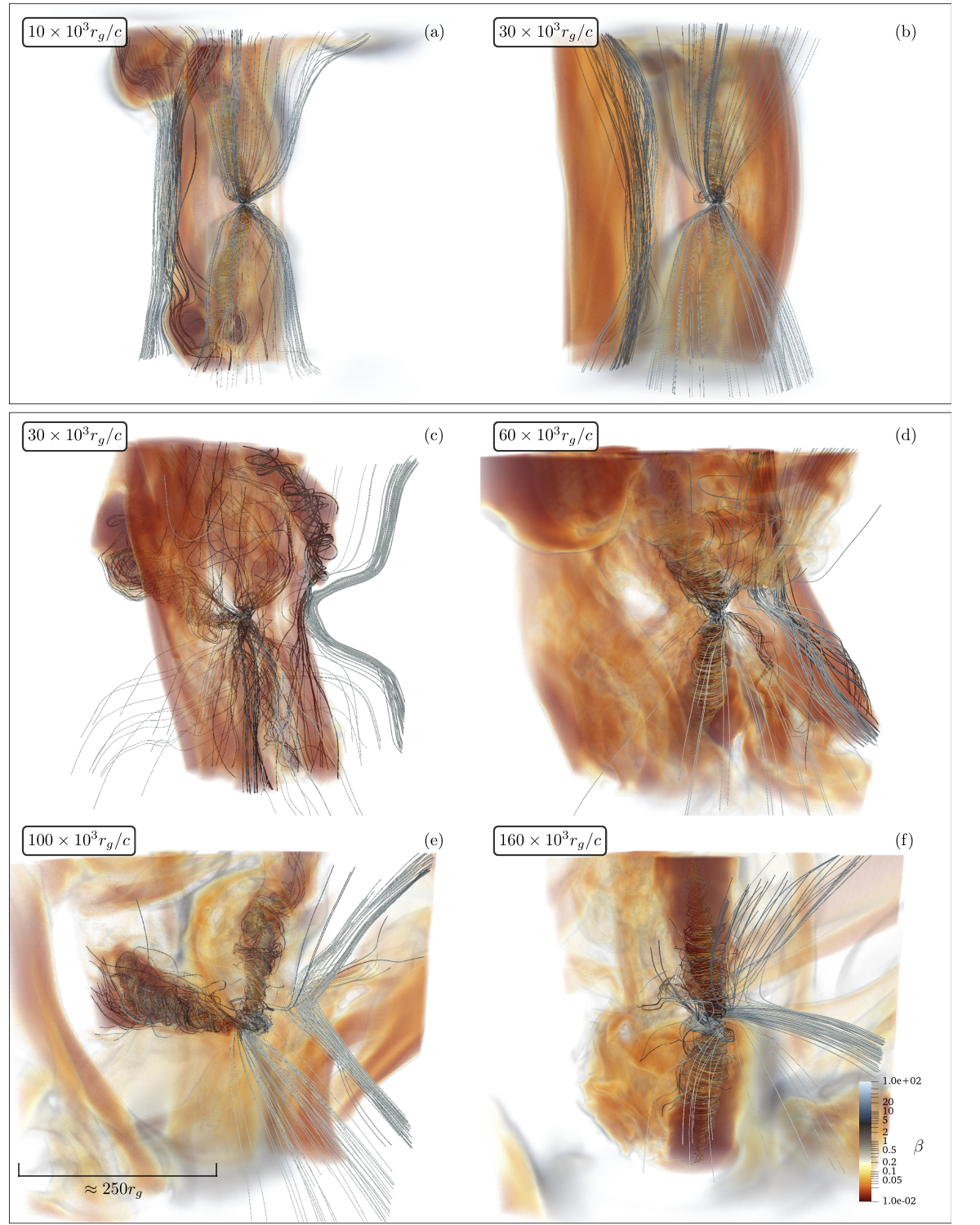}
    \caption{Three-dimensional rendering of the late-time evolution of plasma-$\beta$ (note that the opacity is decreasing as the plasma-$\beta$ is increasing) and magnetic field lines, plotted on a scale of $250r_g$ for $r_b=100r_g$ (top row, (a)-(b)) and $r_b=500r_g$ (bottom, (c)-(f)). The origin of magnetic field lines for rendering is fixed at two regions: around the BH and at $\approx 100r_g$ from the BH. The corresponding movies can be found for \href{https://youtu.be/HxUnv_p7Fj8}{$r_b=100r_g$} and \href{https://youtu.be/YNYKub8YKSs}{$r_b=500r_g$}.
    }
    \label{fig:3d-jet}
\end{figure*}
In Figure~\ref{fig:3d-jet}, we show a three-dimensional rendering of plasma-$\beta$ (in color) and magnetic field lines for $r_b=100r_g$ and $r_b=500r_g$ at $a=0.95$ at a large scale of $500r_g$. The magnetic field lines originate from two regions: on a sphere around the BH and approximately $100r_g$ from the BH. In all snapshots, the spin axis of the BH points upward. The top row ((a)-(b)) shows two snapshots for the case of $r_b=100r_g$. Both demonstrate that the powerful jet extends to large distances and is likely kink-unstable beyond the Bondi radius. Magnetic flux tubes, which are remnants of previous eruption episodes we will discuss in detail in the next section, are present around the BH. Panels ((c)-(f)) show four snapshots for the case of $r_b=500r_g$. At times of $30\times 10^3 r_g/c$ (c) and $100\times10^3 r_g/c$ (e), the jet is weak and is significantly tilted relative to the vertically oriented spin axis of the BH. Between those times, a jet with $\eta\approx 100\%$ reappeared at a time of $60 \times 10^3 r_g/c$ (d). At a time of $160 \times  10^3$ (f), $\phi_{\rm BH}$ increases again (as seen in Fig.~\ref{fig:time-evolution}), leading to a strong jet with $\eta > 100\%$.

\subsubsection{The role of eruptions}
\begin{figure*}
    \centering
    \includegraphics[width=\textwidth]{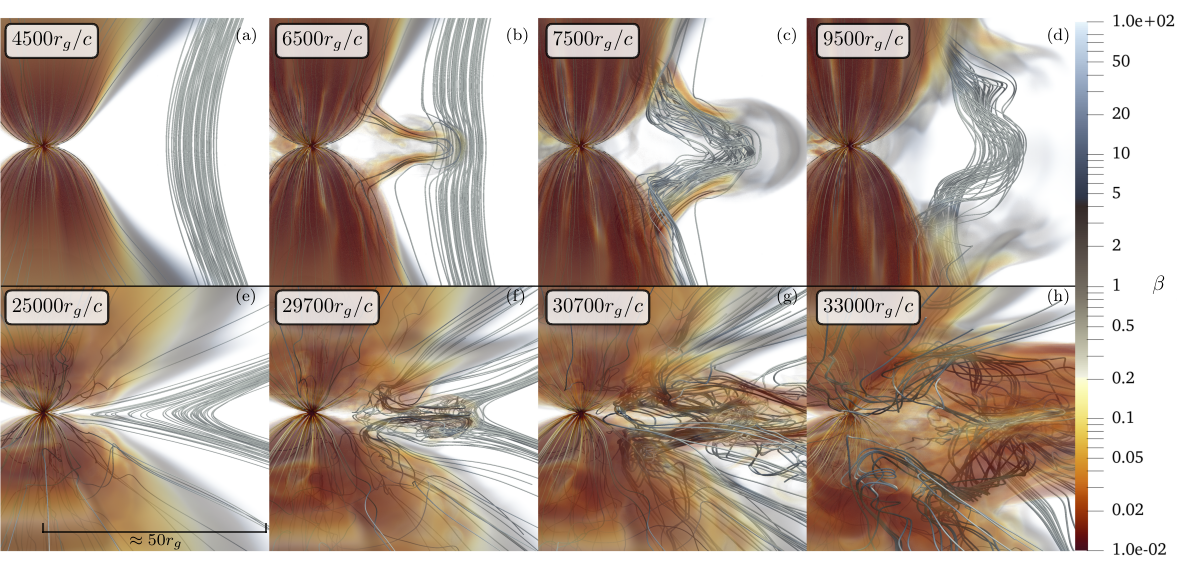}
    \caption{Three-dimensional rendering of the temporal evolution of plasma-$\beta$ (note that the opacity is decreasing as the plasma-$\beta$ is increasing) and magnetic field lines, plotted on a scale of $50r_g$ during the first magnetic flux eruption episode for $r_b=100r_g$ (top row, (a)-(d)) and $r_b=500r_g$ (bottom row, (e)-(h)). Time progresses from left to right. The origin of magnetic field lines for rendering is fixed at two regions: around the BH and at $\approx25r_g$ to the right of the BH in the image plane.}
    \label{fig:3d}
\end{figure*}
We illustrate the distinct dynamics of eruption episodes in simulations with increasing $r_b$ in Figure~\ref{fig:3d}. Similarly to Fig.~\ref{fig:3d-jet}, this figure shows a three-dimensional volume rendering of plasma-$\beta$ (in color) and magnetic field lines for $r_b=100r_g$ (top row, (a)-(d)) and $r_b=500r_g$ (bottom row, (e)-(h)), both for $a=0$. The scale of the plot is approximately $50r_g$. We fix the positions of the starting points of the field lines on a sphere around the BH, as well as at a distance of about $25r_g$ from the BH, along the trajectory of the first eruption. 

Initially, (first column, panels a and e), the system is shown just before the first eruption, at approximately $5000r_g/c$ and $25000r_g/c$ for $r_b=100r_g$ and $r_b=500 r_g$ respectively. Already, there is a noticeable difference in the magnetic field geometry. The case of a smaller Bondi radius exhibits a more vertical field (a), while a larger Bondi radius case shows a more radial field due to the stretching of $B^r$ starting at a greater distance (e).

When the first magnetic flux eruption occurs, it distorts the magnetic field lines ((b) and (f)). Figure~\ref{fig:3d} shows that this distortion is much more dramatic for $r_b = 500 r_g$ than for $r_b = 100 r_g$ (compare panels (g) and (h) to (c) and (d)). Our interpretation is that the smaller $r_b$ case has an unphysically high magnetic pressure at $r \sim 50-100 r_g$ (relative to a more realistic larger $r_b$ case) which suppresses the outward propagation of the eruption. This in turn suppresses how efficiently the eruption can drive turbulence in the ambient medium and reorganize the inflowing magnetic field. In the more realistic larger $r_b$ case (bottom row in Fig.~\ref{fig:3d}) the BH-feedback-driven turbulence changes the magnetic field structure much more dramatically on scales of $\sim 100 r_g$. We show below that it can even reverse the sign of the accreted magnetic flux.



\begin{figure}
    \centering
    \includegraphics[width=\columnwidth]{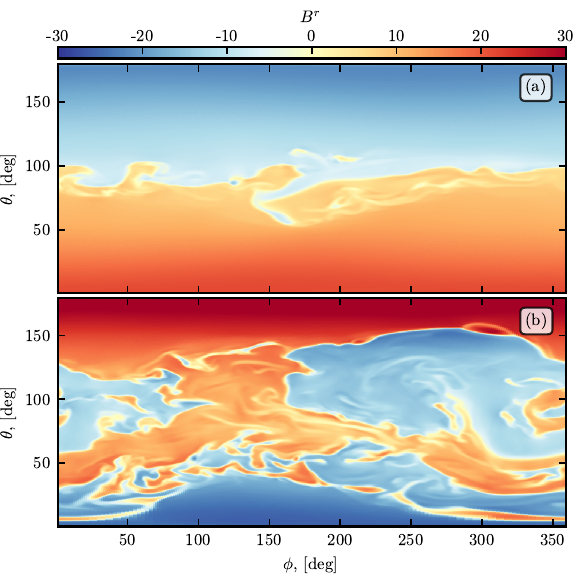}
    \caption{$B^r$ measured on a sphere at the horizon at a time of $30000r_g/c$ for $r_b=100r_g$ (a) and $r_b=500r_g$ (b), both for $a=0.95$.}
    \label{fig:Br}
\end{figure}
As a result of these eruption episodes, the accreting magnetic field is turbulent and less coherent for $r_b=500r_g$, compared to the MAD-like small Bondi radius case. To demonstrate this and its impact on the jet formation, we show the radial magnetic field component, $B^r$, measured on a sphere at the horizon for $a=0.95$ in Figure~\ref{fig:Br}. {These correspond to panels (b) and (c) in Fig.~\ref{fig:3d-jet}.} In a smaller Bondi radius of $100r_g$ case (a), we observe a pattern with two distinct coherent regions: one is characterized by $B^r>0$ in the northern hemisphere ($\theta<90^{\circ}$), and another with $B^r<0$ in the southern hemisphere ($\theta > 90^{\circ}$). In contrast, panel (b) for the larger Bondi radius, $500r_g$, shows the simultaneous existence of multiple regions with different field polarities. This complex polarity of the accreting magnetic field results in less efficient flux accumulation and jet production.

\begin{figure}
    \centering
    \includegraphics[width=\columnwidth]{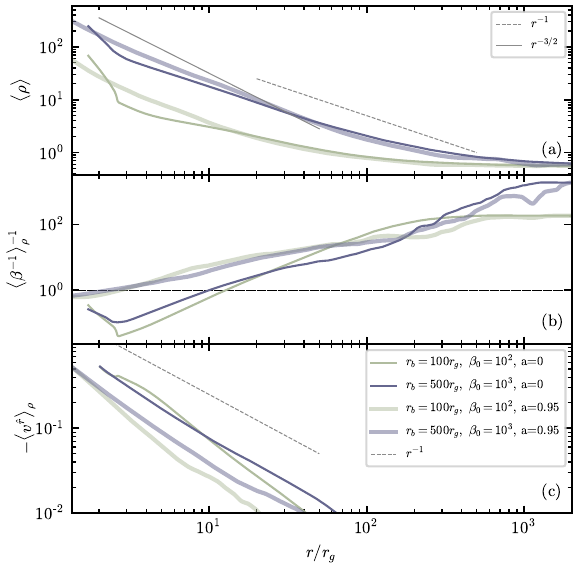}
    \caption{Mean profiles (averaged over a spherical shell and over five snapshots over time of $1000r_g/c$) of density (a), density-weighted plasma-$\beta$ (b), and radial physical velocity (c). The time frames are: $[39-40]\times10^3$ for $r_b=100r_g$ and $a=0.95$; $[39-40]\times10^3$ for $r_b=500r_g$ and $a=0.95$; $[35-36]\times10^3$ for $r_b=100r_g$ and $a=0$; $[100-101]\times10^3$ for $r_b=500r_g$ and $a=0$.}
    \label{fig:profiles}
\end{figure}
We also demonstrate profiles of mean quantities of the accretion flow in Figure~\ref{fig:profiles}, which are averaged over spherical shells at each radius (where $\sigma<10$ to avoid the jet region; throughout the paper, $\sigma=B^2/4 \pi \rho c^2$ denotes the plasma's magnetization) over 10 snapshots spanning $1000r_g/c$ during a period of steady $\phi_{\rm BH}$. These profiles illustrate density (a), density-weighted plasma-$\beta$ (b), and density-weighted radial velocity (c) computed on the tetrad basis (see Appendix~\ref{ap:LNRF}). Similar to Fig.~\ref{fig:time-evolution}, different colors represent different $r_b$, while thin and thick lines represent cases where $a=0$ and $a=0.95$, respectively. The density profiles are all shallower than the Bondi profile, $\rho \sim r^{-3/2}$, which is roughly consistent with $r^{-1}$ as reported in \citet{Ressler2021MNRAS}. The density is higher close to the BH for $r_b=500r_g$ because we begin each simulation with the identical $\rho_0=1$. 

In all cases, we observe that the plasma-$\beta$ value drops below $1$ near the horizon as shown in panel (b), where $\beta\equiv 1$ is represented by the dotted gray line. For the spinning scenarios, plasma-$\beta$ is generally lower, reaching unity, $\beta \approx 1$, within approximately $15 r_g$. However, the profiles are similar across different Bondi radii for a given spin, implying a similar level of magnetic dominance near the BH. The radial velocities follow a similar dependence (c). We return to the evolution of the angular velocity in \S~\ref{sec:mom}.

\subsubsection{Magnetic field evolution}

Since the difference between larger and smaller Bondi radii cases is governed by the presence of magnetized turbulence, driven by eruption episodes, we assess the magnetic field evolution and transport in this section.

\begin{figure}
    \centering
    \includegraphics[width=\columnwidth]{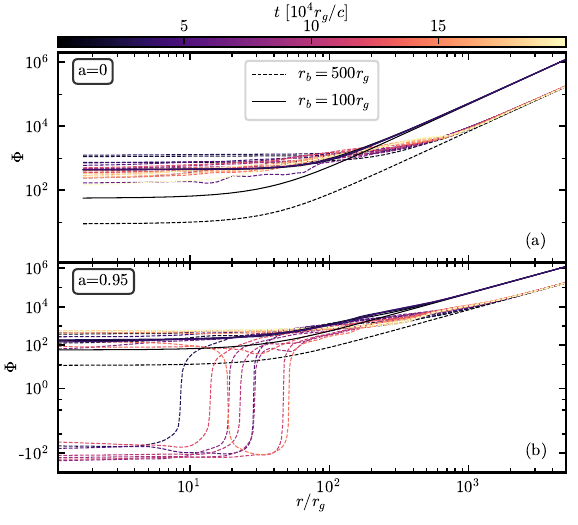}
    \caption{Evolution of magnetic flux, $\Phi_{\rm BH} = \int B^r dA$, profiles (integrated over the northern part of the sphere) for $a=0$ (a) and $a=0.95$ (b) for $r_b=500r_g$ (dashed lines) and $r_b=100r_g$ (solid lines).}
    \label{fig:phi-time}
\end{figure}
First, the difference in the magnetic field transport is reflected in the radial distribution of the magnetic flux. We show the profiles of $\Phi = \int B^r dA$, integrated over the northern hemisphere for $a=0$ (a) and $a=0.95$ (b) in Figure~\ref{fig:phi-time}. Bondi radii of $100r_g$ and $500r_g$ are shown by solid and dashed lines, respectively, the lines are color coded based on the time they are measured (with a cadence of 10000$r_g/c$). 

Far from the BH, all profiles closely follow the initial values at all times, the $r_b=500$ magnetic flux is thus lower due to higher initial plasma-$\beta$. In the case of $r_b=100 r_g$, we find that the profiles quickly reach a steady state within $20000r_g/c$ for both spin values (solid lines in (a) and (b)). In contrast, $r_b=500r_g$ continues to evolve with time due to the turbulent nature of the accreted magnetic field. The difference is more dramatic at $a=0.95$ (b), where $r_b=500r_g$ also shows sign changes in $\Phi$, indicating changes in the direction of magnetic flux. 

\begin{figure}
    \centering
    \includegraphics[width=\columnwidth]{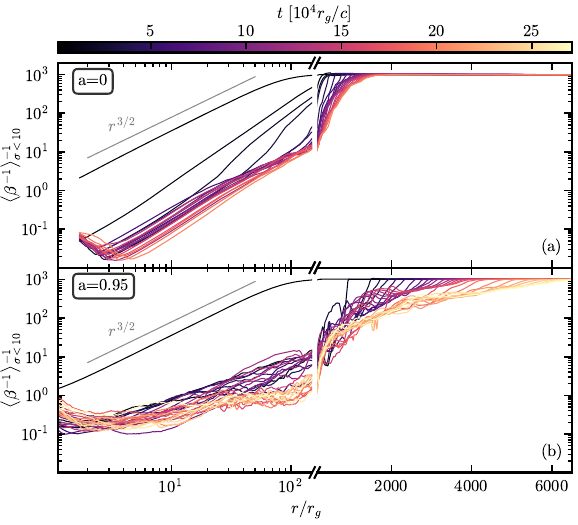}
    \caption{Evolution of plasma-$\beta$ profiles for $r_b=500r_g$ at $a=0$ (a) and $a=0.95$ (b). Early, the profiles follow the expected Bondi profile inside $r_b$ until saturating at low $\beta$ values close to the BH. The low $\beta$ region propagates further from the BH with time.}
    \label{fig:beta-time}
\end{figure}
To further examine the time evolution of the magnetized turbulence, Figure~\ref{fig:beta-time} shows plasma-$\beta$ profiles for $r_b=500r_g$ in a similar way to Fig.~\ref{fig:phi-time}; as before, panels (a) and (b) show the nonspinning and spinning cases, respectively. The averaging is performed over a spherical shell at each radius in the regions where $\sigma<10$ to avoid the jet. The figure is divided into two parts on the horizontal axis: the left side shows a region close to the BH on a logarithmic $r$, while the right side shows linear $r$ at larger radii. 

Initially, both profiles follow $r^{3/2}$ dependence within the Bondi radius, consistent with a Bondi solution with a radial magnetic field. As they evolve, they reach $\beta<1$ inside $r\approx 10r_g$. For a nonspinning BH (a), the $\beta<1$ region extends smoothly to larger distances. For $a=0.95$, instead of steady growth, we observe large fluctuations in the profile due to BH-driven turbulence. The magnetically dominated region of $\beta<1$ can extend up to $100r_g$ at times.

At large distances outside the Bondi radius, the $\beta$ profiles do not saturate -- the low $\beta$ values continue to extend further from the BH in time. In the highly spinning case (b), this extension occurs faster and reaches larger distances. We anticipate that regions of small $\beta$ will continue to extend outward at times later than shown in our simulations.

\subsection{Parameter Dependence}\label{sec:parameters}
\begin{figure*}
    \centering
    \includegraphics[width=\textwidth]{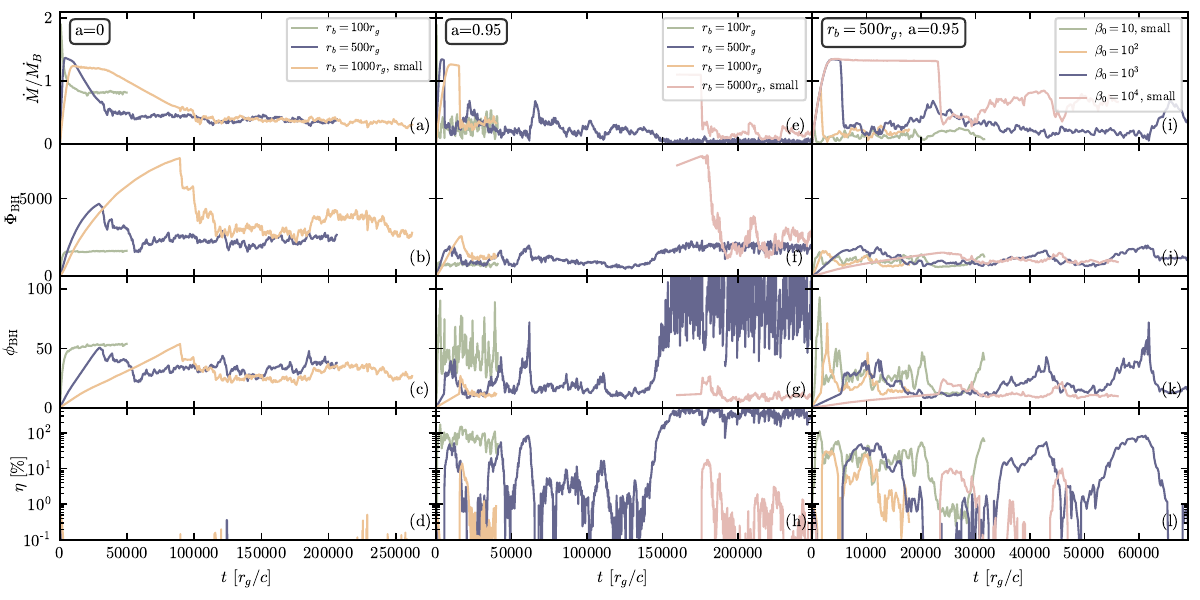}
    \caption{Dependence of temporal evolution of $\dot{M}/\dot{M_B}$, $\Phi_{\rm BH}$, $\phi_{\rm BH}$, and $\eta$ on parameters of our study: Bondi radius at $a=0$ (first column), Bondi radius at $a=0.95$ (second column), and plasma-$\beta$ at $r_b=500$ and $a=0.95$ (third column). All runs are performed at a ``medium'' resolution, except those specified as a ``small'' resolution.}
    \label{fig:time-evolution-params}
\end{figure*}
In this section, we confirm our main findings by expanding our parameter range beyond what is described in the previous section.
The results of all tests are shown in Figure~\ref{fig:time-evolution-params}, structured similarly to Fig.~\ref{fig:time-evolution}. The first row shows the accretion rate $\dot M$, the second row shows the magnetic flux $\Phi_{\rm BH}$, the third row shows the normalized magnetic flux $\phi_{\rm BH}$, and the fourth row shows the energy outflow efficiency $\eta$. Each simulation is represented by a different color. If a simulation was conducted at a lower resolution, it is labeled as ``small'' in the figure.

First, we test the convergence of our results for a given Bondi radius by conducting simulations with $r_b=1000r_g$ for both $a=0$ (at a small resolution, first column) and $a=0.95$ (at a medium resolution, second column) cases. Additionally, we run a simulation with a higher Bondi radius of $5000r_g$. In this case, we extended the outer boundary to $10^5r_g$ and kept the physical resolution inside $10^4r_g$ the same by increasing the number of cells in the radial direction up to 472.

In all simulations with larger Bondi radii, we observe the same main dynamical behavior -- a decrease in the normalized magnetic flux saturation level (panels (c) and (g)), compared to the case of $r_b=100r_g$. Consequently, at $a=0.95$, the outflow energy efficiency decreases, leading to the absence of a powerful jet (panel (h)). Overall, $\dot M$, $\Phi_{\rm BH}$, and $\phi_{\rm BH}$ for $r_b=500r_g$ and $r_b=1000r_g$ reach similar values for both $a=0$ and $a=0.95$. At a higher Bondi radius of $5000r_g$, we find the saturated value of $\phi_{\rm BH}$ to be a little lower, compared to $r_b=1000r_g$.

We also test the sensitivity of our results on the initial $\beta_0$ at $r_b=500r_g$ and $a=0.95$. We conducted a run with $\beta_0=100$ (same as for $r_b=100r_g$) at our medium resolution, and two additional values at a smaller resolution, $\beta_0=10$ and $\beta_0=10^4$ (third column, panels (i)-(l)). Our findings show that the main results converge for $\beta_0=10$ and higher. The magnetic flux is at a similar level (panels (j) and (k)). The energy outflow efficiency is small, indicating that no powerful jet appears. 

For a stronger magnetic field, $\beta_0=10$, we see an oscillatory pattern of $\Phi_{\rm BH}$ while the value of $\dot M$ remains relatively constant, resulting in oscillatory $\phi_{\rm BH}$. This results in the brief emergence of jets with an energy outflow efficiency $\eta\approx 50 \%$, though the time-averaged jet efficiency is again low, $\eta \sim 20 \%$.

\subsection{Late evolution and impact of gas angular momentum}\label{sec:mom}
In this section, we explore the dependence of our results on the final free parameter of our problem -- angular momentum $l$. For the toroidal component of the four-velocity, we define:
\begin{equation}
    u^\phi = \frac{l}{r^2} \sin^2 \theta = \frac{\sqrt{r_c}}{r^2} \sin^2 \theta,
\end{equation}
where $r_c$ is the circularization radius. This approach concentrates the rotation at the equatorial plane, resulting in the physical velocity $u^{\hat{\phi}}$ far from the BH:
\begin{equation}
    u^{\hat{\phi}} = r \sin \theta u^\phi = \frac{l}{r} \sin^3 \theta,
\end{equation}
which provides the physical angular velocity $\Omega$: 
\begin{equation}
    \Omega = u^{\hat{\phi}} / (r \sin \theta) \propto \sin^2 \theta.
\end{equation}
However, we find that the outcomes are insensitive to the exact initial distribution of angular momentum: we have checked the alternative cases of angular velocities, $\Omega \propto \sin^0 \theta$ and $\Omega \propto \sin^1 \theta$.

\begin{figure}
    \centering
    \includegraphics[width=\columnwidth]{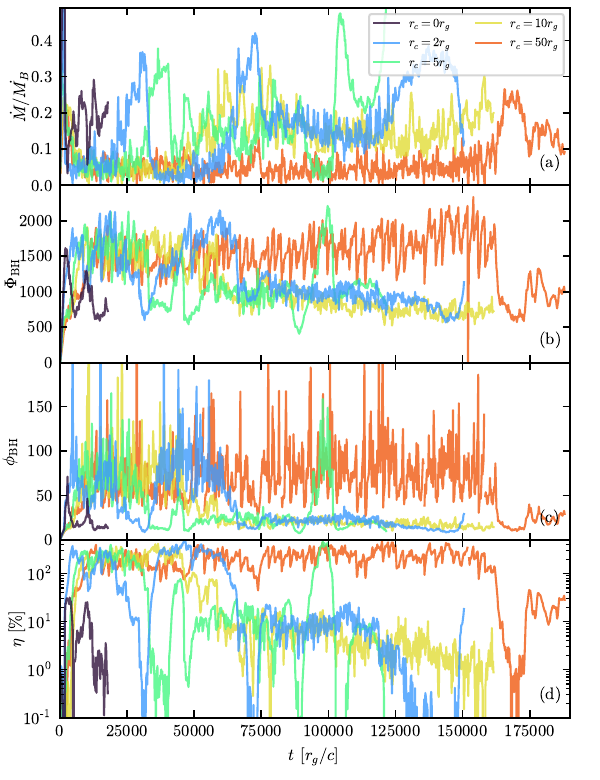}
    \caption{Time evolution of main flow characteristics (organized similarly to Fig.~\ref{fig:time-evolution}) for simulations with $r_b=500r_g$, $\beta_0=10^2$, $a=0.95$ and varying initial circularization radius, $r_c$, represented by different colors. The simulation that started with zero angular momentum in the flow, $r_c=0r_g$, is performed at a ``medium'' resolution, while all simulations with $r_c>0$ are performed at a ``small'' resolution. }
    \label{fig:ang-momentum}
\end{figure}
We fix $r_b=500r_g$ (as our results converge with increasing Bondi radius), $a=0.95$, and $\beta_0=100$ (since our results converge with increasing $\beta_0$). We vary $r_c$ at a small resolution (as our results converge with resolution) from $r_c=2r_g$ (or $0.004r_b$) to $50r_g$ (or $0.1r_b$). The results are shown in Figure~\ref{fig:ang-momentum}, where each $r_c$ is represented by a different color. 

Initially, all simulations reach a high value of $\Phi_{\rm BH}$, similar to the $r_c=0$ case. The larger the $r_c$-value, the longer the period during which $\Phi_{\rm BH}$ remains at a high level. This elevated $\Phi_{\rm BH}$, as well as $\phi_{\rm BH}$, is similar across all $r_c$ values. This state is characterized by low $\dot M$, high efficiency $\eta$ exceeding $100\%$, and large oscillations in $\Phi_{\rm BH}$ -- characteristics typical of the MAD state. 

Eventually, all simulations transition to a state similar to that observed for $r_c=0$, where $\Phi_{\rm BH}$ and $\phi_{\rm BH}$ decrease and $\eta \approx 1\%$. Simultaneously, $\dot M$ increases. Notably, the low levels of $\Phi_{\rm BH}$ and $\phi_{\rm BH}$ are approximately the same across all simulations, regardless of different $r_c$ values. For the lower $r_c$ values (e.g., $r_c=2r_g$ and $5r_g$),  we find a periodic emergence of powerful jets, characterized by approximately the same level of $\Phi_{\rm BH}$ while $\dot M$ increases.

\begin{figure*}
    \centering
    \includegraphics[width=\textwidth]{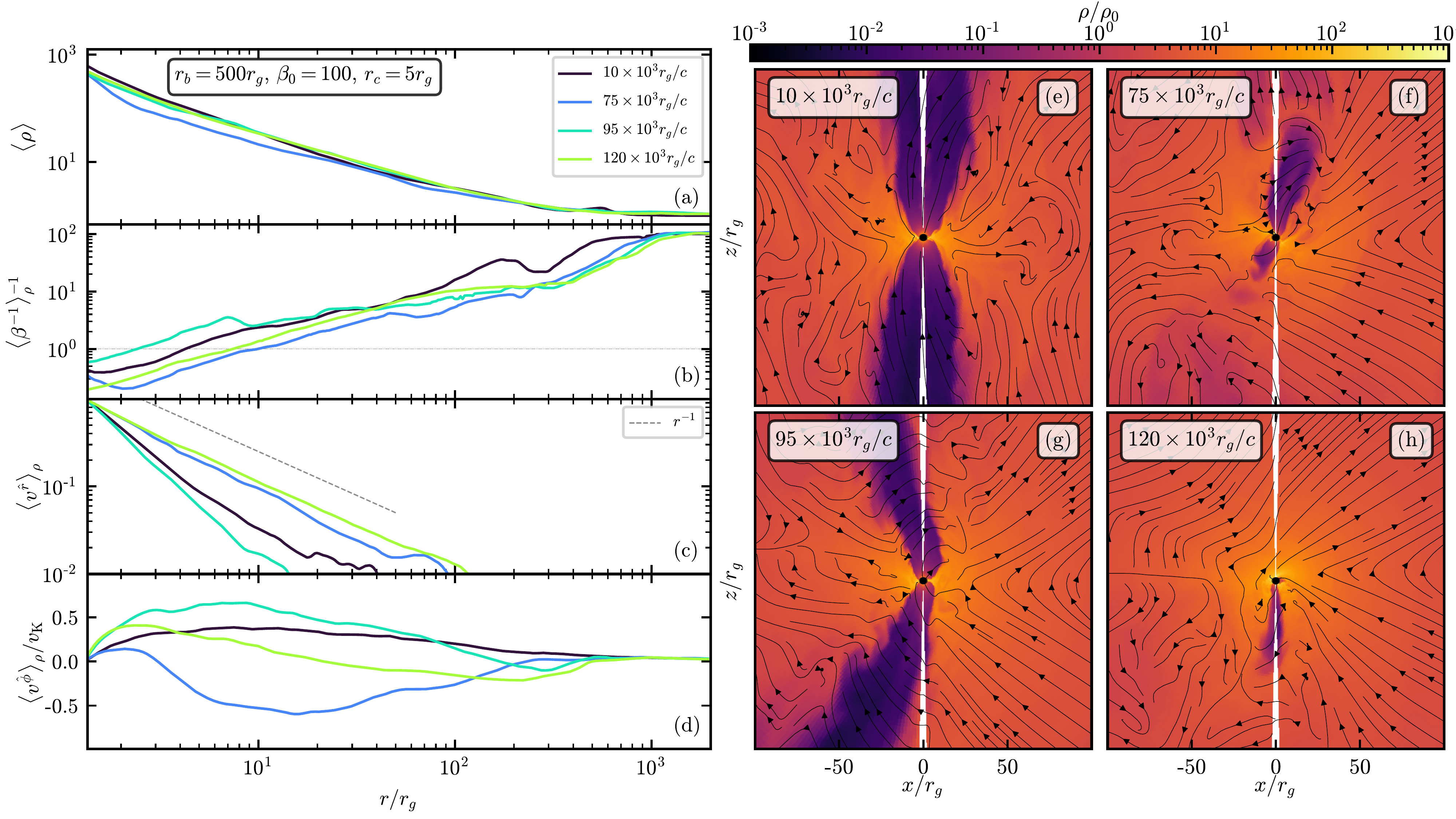}
    \caption{Mean profiles of density (a), $\rho$-weighted plasma-$\beta$ (b), and radial (c) and angular velocities (d) at four distinct times for $r_b=500r_g$, $a=0.95$, and $r_c=5r_g$. Panels ((e)-(h)) show snapshots of density at these times in the poloidal plane; black lines represent the magnetic field lines. The angular velocity is normalized by the Keplerian velocity profile at $\theta=\pi/2$.}
    \label{fig:ang-momentum-5rg}
\end{figure*}
We find that the transitions between high and low $\Phi_{\rm BH}$ (and correspondingly high and low jet efficiency $\eta$) are driven by changes in the angular momentum distribution in the system. This is illustrated in Figure~\ref{fig:ang-momentum-5rg}, where panels ((a)-(d)) show radial profiles of several quantities at four distinct times, and panels ((e)-(h)) show the corresponding snapshots of plasma density in the poloidal plane at these times. Panels ((a)-(d)) demonstrate the profiles of average density (a), $\rho$-weighted plasma-$\beta$ (b), and radial (c) and angular (d) physical velocities. 

At $10\times10^3 r_g/c$, Fig.~\ref{fig:ang-momentum} demonstrates that $r_c=5r_g$ case is in a typical MAD state. The snapshot (Fig.~\ref{fig:ang-momentum-5rg}e) shows that the flow is organized into a thick disk and a powerful jet is present. The angular velocity is positive and high. Later, at $75 \times 10^3 r_g/c$, $\Phi_{\rm BH}$ falls; the inflow becomes less organized, with the jet becoming weaker and tilted to the side (f). At this time, the angular momentum profile shows small positive values near the BH and negative values beyond $\approx 3 r_g$ (d). The jet reappears at $95\times10^3 r_g/c$ (g), which again correlates with high angular momentum (d). Finally, the jet partially disappears at $120\times10^3 r_g/c$ (h), while the angular momentum magnitude decreases again (d). 

Thus, the periods of high and low $\Phi_{\rm BH}$ (and, consequently, the periods of high and low $\eta$ resulting in the presence and absence of powerful jets) are closely correlated with the evolution of angular momentum in the accreting plasma. We attribute this difference to the influence of angular momentum on the dynamical behavior of eruptions. When the angular momentum is high, the erupting material is quickly sheared and decorrelated by the rotation. In contrast, when the angular momentum is low, we observe behavior similar to that in Fig.~\ref{fig:3d}, where eruptions inhibit magnetic flux accumulation. Additionally, high angular velocity (d) is correlated with low radial velocity (c) and vice versa. This highlights the decreased inflow and accretion when there is significant angular momentum.

One perhaps surprising result of Figure \ref{fig:ang-momentum-5rg} is that the angular momentum is not solely described by its initial distribution but can also self-consistently evolve over long timescales. We find a correlation between angular momentum profiles and plasma-$\beta$ profiles: when rotation decreases (Fig.\ref{fig:ang-momentum-5rg}d), the state is characterized by lower values of plasma-$\beta$ (Fig.\ref{fig:ang-momentum-5rg}b) and vice versa. This could indicate increased torques from strong magnetic fields, which remove angular momentum from the flow. Further study is required to validate this conjecture. 

\begin{figure}
    \centering
    \includegraphics[width=\columnwidth]{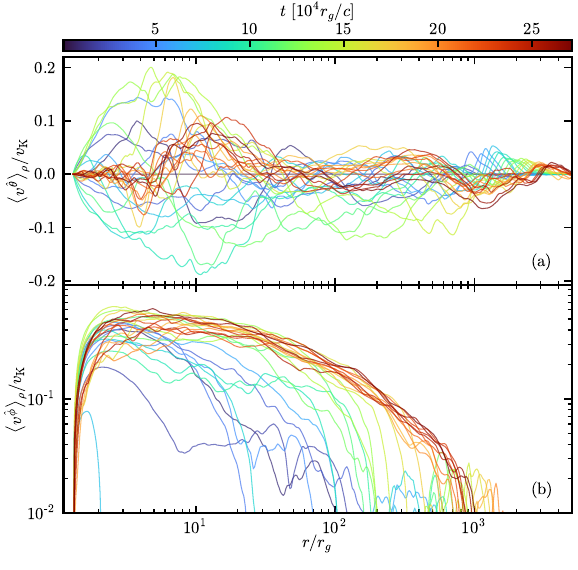}
    \caption{Time evolution of $\rho$-weighted profiles of physical components of the gas velocity, $\langle v^{\hat \theta} \rangle_\rho$, (a); and, $\langle v^{\hat \phi} \rangle_\rho$, (b) measured in the LNRF, for the case of $r_b=500r_g$, $\beta_0=10^3$, $r_c=0$, $a=0.95$. The velocities are normalized by the Keplerian velocity in the LNRF at $\theta=\pi/2$.}
    \label{fig:angular-v-time}
\end{figure}
When the BH is spinning, $a=0.95$, the angular momentum of the accreting material evolves over time even in the absence of initial angular momentum, $r_c=0$. In Figure~\ref{fig:angular-v-time}, we show the evolution of the $\rho$-weighted profiles of physical velocities $v^{\hat \theta}$ (a) and $v^{\hat \phi}$ (b), normalized by the Keplerian velocity profile at $\theta=\pi/2$, computed in a locally nonrotating frame (LNRF; see Appendix~\ref{ap:LNRF}). This simulation ($r_b=500r_g$ and $\beta_0=10^3$), was described in detail in \S~\ref{sec:main}. The rotation is typically, but not always, aligned with the spin axis of the BH, e.g., when $v^{\hat \phi} \approx 0.2 v_{\rm K} $ at $t\approx \times 10 \times 10^4r_g/c$, $v^{\hat \theta} \approx -0.2 v_{\rm K}$. Notably, we find that the powerful jet with $\eta\approx 100\%$ (as shown in Fig.~\ref{fig:time-evolution}) is associated with an increased angular velocity aligned with the spin axis of the BH. In particular, the jet appears at $t\approx 160 \times 10^3 r_g/c$, which corresponds to the high-angular-momentum state shown by the red lines with $v^{\hat \phi}\approx 0.5 v_{\rm K}$ (b). As in our simulations with initial angular momentum, we expect the gas to eventually return to a low-angular-momentum phase, leading to the subsequent absence of a powerful jet. The long-term duty cycle of phases with strong jets is unclear.

\section{Conclusion and discussion}
Using ideal GRMHD simulations, we study spherical accretion of plasma with low angular momentum and net magnetic flux. We find that this configuration leads to a magnetically dominated plasma close to the black hole, characterized by an average plasma-$\beta \leq 1$. However, despite an abundance of magnetic flux available in the system and the strong magnetization close to the black hole, some key features of MADs \citep{Tchekhovskoy2011} are not present. Notably, the dimensionless magnetic flux on the BH horizon saturates below the nominal value of $\phi_{\rm BH}\approx 50$ typically observed in the MAD state. As a result, for a spinning BH, the system only produces a weak jet.

We attribute this significant departure from the typical MAD state to the effect of turbulence driven in the ambient medium by horizon-scale flow dynamics. As magnetic flux at the horizon accumulates, it leads to magnetic reconnection events that drive bubbles of magnetized plasma out into the accretion flow. This turbulence suppresses the accretion of magnetic flux leading to a weaker jet. In comparison to the standard MAD state, we do not find large fluctuations in the magnetic flux $\Phi_{\rm BH}$ at the horizon, indicating that large-scale magnetic eruptions are not present in the saturated state. There are, however, continued small eruption episodes that drive turbulence in the ambient medium, as indicated by smaller-scale fluctuations in the horizon-scale magnetic flux {(Figures~\ref{fig:time-evolution}(f)-(g))}. Interestingly, we find that $\phi_{\rm BH}$ and the jet power are roughly independent of the Bondi radius for  $r_b = 500r_g$ and above. The initial spherical GRMHD simulations of \citet{Ressler2020ApJ...896L...6R} focused on $r_b = 200 r_g$ and found larger jet power and horizon magnetic flux (as we do here for  $r_b = 100 r_g$); we attribute this to the higher equatorial pressure at the (small) Bondi radius artificially suppressing the effect of BH-feedback-driven turbulence on magnetic flux accumulation.

In our interpretation, turbulence driven by episodic eruptions of magnetic flux accumulated on the black hole likely enhances the turbulent transport of magnetic flux outward, which counteracts the laminar advection of magnetic flux inward. Note that the inflow speed is much smaller than the freefall velocity (Figure~\ref{fig:profiles}(c)) so modest levels of turbulence in the ambient medium can suppress the accumulation of magnetic flux on the black hole. It is unclear, however, why the net result is a dimensionless magnetic flux of $\phi \sim 10$ (for $a = 0.95$) over a range of parameters. A mean-field model of the competition between radial advection and eruption-driven transport of magnetic flux could be valuable for understanding this better.\footnote{Standard accretion disk models of flux transport (e.g., \citealt{Lubow1994}) are not applicable to the low-angular-momentum case considered here.}

To understand the transition between the initially nonrotating plasma simulated here and the rotating torus initial conditions that have been the focus of previous work (e.g., \citealt{Tchekhovskoy2011}), we carried out a series of simulations with increasing angular momentum, as parameterized by the circularization radius $r_c$ of the inflowing plasma (all in the regime of $r_c < r_b$). Finite angular momentum initially leads to more efficient horizon-scale magnetic flux accumulation and a stronger jet. We argue that rotation of the plasma efficiently shears out and decorrelates the magnetic eruptions, significantly suppressing the extent of the turbulence in the flow and thus enhancing the accumulation of magnetic flux onto the BH.

One of the more striking results of our simulations is that the rotation of the plasma in the inner $\sim 10-100 r_g$ undergoes secular evolution on long timescales, independent of whether the plasma is initially rotating or nonrotating. Simulations with and without initial rotation of the flow both experience phases of low and high angular momentum at later times.  

For initially nonrotating plasma accreting onto a spinning BH, a powerful jet, $\eta>100\%$, can emerge for short periods of time. These episodes are correlated with an increase in the angular momentum of the accreting plasma aligned with the spin axis of the BH. We argue that this angular momentum is due to a combination of angular momentum in the near-horizon region (due to the spinning BH) transferred to large radii by magnetic torques and random angular momentum associated with the turbulence driven in the ambient medium.

For initially rotating plasma accreting onto a spinning BH, we find that the dimensionless horizon-scale magnetic flux, $\phi_{\rm BH}$, can decrease for extended periods of time, at which point the strong jet also disappears. This is true for all values of the initial circularization radius that we simulated, though the jet only disappears at late times, $\sim 170,000r_g/c$, in our largest $r_c = 50 r_g$ simulation. The decrease in $\phi_{\rm BH}$ and jet power are strongly correlated with a decrease in the angular momentum of the near-horizon plasma. The loss of angular momentum may be due to the torques exerted by the magnetic field in the magnetically dominated low-$\beta$ plasma generically present close to the BH.

Our results indicate that both the angular momentum content and horizon-scale magnetic flux of radiatively inefficient accretion flows can evolve much more in time than previously appreciated. Moreover, we have shown that these two dynamical components are in fact intimately related: in our simulations, large dimensionless magnetic flux and powerful jets appear only when the plasma in the inner $\sim 10-100 r_g$ has significant angular momentum (even though the plasma close to the BH is strongly magnetized with $\beta \lesssim 1$ in all cases we have studied). The angular momentum content of the plasma is not, however, just a property of the ``initial condition'' at large radii, but dynamically evolves due to strong magnetic torques and turbulence driven in the surrounding medium by jets and near-horizon magnetic eruptions. It is not at all clear what the long-timescale statistical steady state for any of our simulations actually is, whether it is captured by the simulations we have carried out, or whether the angular momentum, magnetic flux, and jet properties will continue to secularly evolve. It is also striking that the plasma continues to get more strongly magnetized at most radii as a function of time as indicated by the evolving $\beta$ profiles shown in Figure \ref{fig:beta-time}. More work is clearly needed to understand the very long-timescale evolution of these systems.

We have restricted our study to an initial magnetic field in the laboratory frame that is strictly poloidal and aligned with the spin axis of the BH. However, simulations with a Bondi radius of $200r_g$ and zero angular momentum \citep{Ressler2021MNRAS} have shown that tilting the magnetic field affects the resulting jet power. It remains unclear whether this carries over to the larger Bondi radius regime of most astrophysical interest. Therefore, in future studies, it would be interesting to explore the impact of tilting the magnetic field, as well as gas rotation, relative to the spin axis of the BH. These model problems could help in understanding the connection between the finite circularization radius problem studied here and simulations that begin with an equilibrium torus initial condition, in which tilting the magnetic field and plasma rotation axis relative to the BH spin axis can dramatically change the dynamics \citep[e.g.,][]{Liska2018MNRAS.474L..81L,Liska2019MNRAS.487..550L}. Inflowing gas with a finite circularization radius likely serves as a better initial condition for the accretion of hot gas in the centers of galaxies than an equilibrium torus \citep[e.g.,][]{Cuadra2006MNRAS.366..358C, Guo2024ApJ}.

The results presented here on the interplay between accretion and jets in GRMHD simulations of spherical accretion build on those of \citet{Ressler2021MNRAS}, \citet{Kwan2023ApJl}, and \citet{Lalakos2024ApJ}; we have extended these author's results by varying the black hole spin,  the initial gas angular momentum, the Bondi radius, and the initial plasma-$\beta$.  For zero gas angular momentum, we find a similar dynamical evolution to that reported by \citet{Lalakos2024ApJ} -- the initial emergence of a powerful jet with $\eta>100\%$, followed by its eventual disappearance as the normalized magnetic flux at the horizon ($\phi_{\rm BH}$) saturates below the typical MAD value. However, we attribute this behavior to the turbulence driven by eruption episodes, rather than the jet's current-driven kink instability, as proposed by \citet{Lalakos2024ApJ}. This interpretation is supported by our investigation of the $a=0$ case, where no jets are produced. Even in the absence of jets, the coherence of the accreted magnetic field is disturbed, leading to less efficient magnetic flux accumulation. In addition, if the transitions from $\eta > 100 \%$ to a low $\eta$ state were driven by jet feedback on large scales, we would expect the transition to take longer for a larger Bondi radius and for the accumulated magnetic flux to plausibly depend on the Bondi radius.  Neither of these is true in our simulations.  For example, our results show that the duration of the high-$\eta$ jet phase depends weakly on the Bondi radius, as seen in the initial spike in $\eta$ in Fig.~\ref{fig:time-evolution-params}. This is consistent with the disruption of magnetic flux accumulation originating in feedback from magnetic eruptions close to the black hole, rather than from larger radii. Once the strong jet is disrupted, however, the remaining weaker jets are indeed more prone to kink instability, which generates weak jets that dissipate their energy over a large solid angle (Fig. \ref{fig:3d-jet}).

The subtlety of magnetic flux accumulation even in the simplest problem of spherical accretion onto a BH also raises the possibility that the presence of a strong jet may depend on other quantities that influence the magnetic field evolution and its topology. This includes, e.g., the difference in magnetic reconnection rates in fluid and collisionless plasmas. The initial GRPIC simulations of spherical accretion by \citet{Alisa2023} did not find a difference in horizon-scale magnetic flux relative to fluid simulations, but it is important to note that these simulations only studied the initial burst of accretion at early times, which is not necessarily representative of the long-timescale magnetic flux (compare, e.g., early and late times in Fig.~\ref{fig:time-evolution}).   

One possible application of our findings is to accretion onto Sgr A*. In their simulations of accretion onto Sgr A* from the surrounding stellar winds, \citet{Ressler2023MNRAS.521.4277R} found periods of time with large dimensionless magnetic flux and strong jets, and periods of time with weak jets. These may be analogous to the high- and low-angular-momentum states found here. More broadly, our results suggest that in the low-angular-momentum state explored in this work, observations should indicate the presence of strong magnetic fields even in the absence of a powerful jet. This is potentially consistent with the EHT and GRAVITY evidence for strong magnetic fields in Sgr A* \citep{GRAVITYA&A2018, Gravity2023A&A, EHTsgr2024ApJ...964L..26E}, but the lack of definitive evidence for a powerful jet. As a check on this interpretation, we performed ray-tracing calculations of the synchrotron emission in the strong and weak jet states found in this work (as described in \S~\ref{sec:main}). We find that the linear and circular polarization levels are indeed similar in both cases. Low angular momentum in the inflowing gas could also help explain the surprising lack of azimuthal asymmetry in the EHT image of Sgr A* \citep{Faggert24}.

\begin{acknowledgments}
A.G. and A.P. thank Misha Medvedev for fruitful discussions; we also thank Sasha Tchekhovskoy for useful conversations.  The simulations presented in this work were performed on computational resources managed and supported by Princeton Research Computing, a consortium of groups including the Princeton Institute for Computational Science and Engineering (PICSciE) and the Office of Information Technology's High Performance Computing Center and Visualization Laboratory at Princeton University. A.P. is supported by a grant from the Simons Foundation (MP-SCMPS-00001470). A.P. also acknowledges support from NASA grant 80NSSC22K1054, NSF grants PHY-2231698 and PHY-2206610, by an Alfred P.~Sloan Research Fellowship and a Packard Foundation Fellowship in Science and Engineering. M.L. was supported by the NASA Hubble Fellowship Program award HST-HF2-51537.002-A and NASA Astrophysics Theory Program award 80NSSC22K0817. This research benefited from interactions at the Kavli Institute for Theoretical Physics, supported by NSF PHY-2309135. 
\end{acknowledgments}

\appendix

\section{Numerical convergence}\label{ap:convergence}
\begin{figure*}
    \centering
    \includegraphics[width=0.5\textwidth]{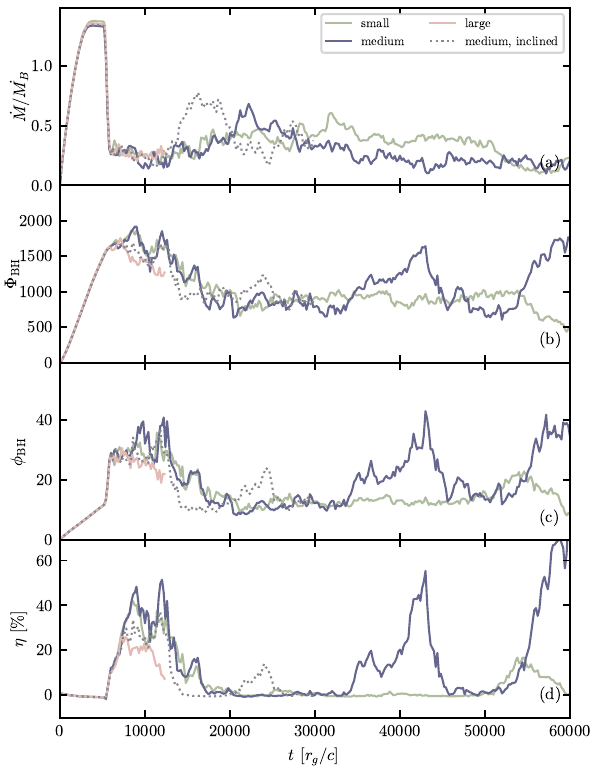}
    \caption{Comparison of the time evolution of the accretion rate $\dot M$ (a), magnetic flux at the horizon $\Phi_{\rm BH}$ (b), normalized dimensionless magnetic flux at the horizon $\phi_{\rm BH}$ (c), and efficiency $\eta$ (d) for the same accretion problem at different resolutions, represented by different colors. The dotted lines demonstrate simulation at a ``medium'' resolution performed by inclining the system by $90 \deg$. }
    \label{fig:convergence}
\end{figure*}
We perform the main simulation, $r_b=500r_g$, $\beta_0=10^3$, and $a=0.95$ at three different resolutions: small, medium (shown in the main text), and large. These correspond to grid sizes $384 \times 144 \times 256$, $812 \times 384 \times 512$, and $1536 \times 768 \times 1024$ in $N_r \times N_\theta \times N_\phi$. The time evolution of $\dot{M}$, $\Phi_{\rm BH}$, $\phi_{\rm BH}$, and $\eta$ are shown in Figures~\ref{fig:convergence}((a)-(d)). Both small and medium resolutions show convergence with time. Due to high computational cost, we run the large resolution simulation to the point where $\phi_{\rm BH}$ falls, converging with other resolutions. Another uncertainty could arise due to the spherical grid's polar singularity. We incline the BH spin by $90 \deg$ at medium resolution to test this. Therefore, the jet launches perpendicular to the polar singularity. We see the convergence of main parameters relative to the grid inclination (Figures~\ref{fig:convergence}(a)-(d), dashed and solid purple lines).

\section{Tetrad transformation}\label{ap:LNRF}
We utilize widely used notations in a Boyer-Lindquist (BL) coordinates $t, r , \theta, \phi$ ($M\equiv c \equiv 1$):
\begin{equation}
    \Delta \equiv r^2  - 2 r + a^2, \ \Sigma \equiv r^2 + a^2 \cos^2 \theta, \ A \equiv (r^2 + a^2)^2 - a^2 \Delta \sin^2 \theta.
\end{equation}
Covariant basis vectors of the LNRF in BL coordinates \citep{Bardeen1972ApJ...178..347B}:
\begin{equation}
    \begin{split}
        e^{\hat t}_{\mu} &= \sqrt{\frac{\Sigma \Delta}{A}} \delta^t_{\mu}, \\
        e^{\hat r}_{\mu} &= \sqrt{\frac{\Sigma}{\Delta}} \delta^{r}_{\mu}, \\
        e^{\hat \theta}_{\mu} &= \sqrt{\Sigma} \delta^\theta_\mu, \\
        e^{\hat \phi}_{\mu} &= -\frac{2 a r \sin \theta}{\sqrt{\Sigma A}} \delta^t_\mu + \sqrt{\frac{A}{\Sigma}} \sin \theta \delta^\phi_\mu.
    \end{split}
\end{equation}
Transforming these basis vectors to KS coordinates $\tilde \mu$ ($r$ and $\theta$ do not change from BL to KS):
\begin{equation}
    \begin{split}
        e^{\hat t}_{\tilde{\mu}} &= \sqrt{\frac{\Sigma \Delta}{A}} \delta^{\tilde t}_{\tilde \mu} - \frac{2r }{\Delta}\sqrt{\frac{\Sigma \Delta}{A}} \delta^{\tilde r}_{\tilde \mu}, \\
        e^{\hat r}_{\tilde \mu} &= \sqrt{\frac{\Sigma}{\Delta}} \delta^{\tilde r}_{\mu}, \\
        e^{\hat \theta}_{\tilde \mu} &= \sqrt{\Sigma} \delta^{\tilde \theta}_\mu, \\
        e^{\hat \phi}_{\tilde \mu} &= -\frac{2 a r \sin \theta}{\sqrt{\Sigma A}} \delta^{\tilde t}_{\tilde \mu} + \left( \frac{4 a r^2 \sin \theta }{\Delta \sqrt{A \Sigma}} - \frac{a \sin \theta}{\Delta} \sqrt{\frac{A}{\Sigma}} \right)\delta^{\tilde r}_{\tilde \mu}  + \sqrt{\frac{A}{\Sigma}} \sin \theta \delta^{\tilde \phi}_{\tilde \mu}.
    \end{split}
\end{equation}
Finally, three-velocity in tetrad basis of BL coordinates, transformed from four-velocity $u^{\tilde \mu}$ in KS coordinates:
\begin{equation}
    v^{\hat i} = \frac{e^{\hat i}_{\tilde \mu} u^{\tilde \mu}}{e^{\hat t}_{\tilde \mu} u^{\tilde \mu} }.
\end{equation}

The Keplerian four-velocity in BL coordinates is $u_{\rm K}^{\mu} =u^t (\delta_t^{\mu}  + \Omega_{\rm K} \delta_\phi^\mu)$, where $\Omega_{\rm K} = (a + r^{3/2})^{-1}$. Using the tetrad basis for transforming the Keplerian contravariant velocity in BL coordinates to LNRF:
\begin{equation}
    v_{\rm K}^{\hat \phi} = \frac{u^{\hat \phi}_{\rm K}}{u^{\hat t}_{\rm K}} = - \frac{2 a r \sin \theta}{\Sigma \sqrt{\Delta}} + \frac{A}{\Sigma \sqrt{\Delta}} \sin \theta \Omega_{\rm K} = \frac{\sin \theta}{\Sigma \sqrt{\Delta}} \left( A \Omega_{\rm K} - 2 a r \right).
\end{equation}
We thus normalize velocities in LNRF by Keplerian velocity in the equatorial plane $v_{\rm K} \equiv v_{\rm K}^{\hat \phi}(\theta= \pi/2)$.

\bibliography{main}{}
\bibliographystyle{aasjournal}

\end{document}